\documentclass[10pt,twocolumn,british,tightenlines,eqsecnum,floats,aps,amsmath,amssymb,nofootinbib,superscriptaddress,prd,showpacs,showkeys]%
{revtex4}
\usepackage{array}
\usepackage{booktabs}
\usepackage{tabu}
\usepackage{dcolumn}
\usepackage{subfigure}
\usepackage{graphicx}% Include figure files
\usepackage{dcolumn}% Align table columns on decimal point
\usepackage{bm}% bold math
\usepackage[latin9]{inputenc}
\usepackage{color}
\usepackage[normalem]{ulem}
\usepackage{babel}
\usepackage{amsmath}
\usepackage{amssymb}
\usepackage{graphicx}
\usepackage{esint}
\usepackage[unicode=true,pdfusetitle,
 bookmarks=true,bookmarksnumbered=false,bookmarksopen=false,
 breaklinks=false,pdfborder={0 0 1},backref=false,colorlinks=false]
 {hyperref}

\makeatletter
\@ifundefined{textcolor}{}
{%
 \definecolor{BLACK}{gray}{0}
 \definecolor{WHITE}{gray}{1}
 \definecolor{RED}{rgb}{1,0,0}
 \definecolor{GREEN}{rgb}{0,1,0}
 \definecolor{BLUE}{rgb}{0,0,1}
 \definecolor{CYAN}{cmyk}{1,0,0,0}
 \definecolor{MAGENTA}{cmyk}{0,1,0,0}
 \definecolor{YELLOW}{cmyk}{0,0,1,0}
}
\usepackage{color}\usepackage{euscript}\usepackage{epsfig}
\usepackage{amsfonts}\usepackage{fancyhdr}
\@ifundefined{textcolor}{}
{%
 \definecolor{BLACK}{gray}{0}
 \definecolor{WHITE}{gray}{1}
 \definecolor{RED}{rgb}{1,0,0}
 \definecolor{GREEN}{rgb}{0,1,0}
 \definecolor{BLUE}{rgb}{0,0,1}
 \definecolor{CYAN}{cmyk}{1,0,0,0}
 \definecolor{MAGENTA}{cmyk}{0,1,0,0}
 \definecolor{YELLOW}{cmyk}{0,0,1,0}
}

\makeatother

\begin{document}

\title{Cosmic Acceleration and Geodesic Deviation in Chameleon Scalar Field Model}

\author{Raziyeh Zaregonbadi}
  \email{r.zaregonbadi@malayeru.ac.ir}
  \affiliation{Department of Physics, Faculty of Science, Malayer University, Malayer, Iran}

\author{Nasim Saba}
 \email{n_saba@sbu.ac.ir}
 \affiliation{Department of Physics, Shahid Beheshti University, Evin, Tehran, 19839, Iran}

\author{Mehrdad Farhoudi}
 \email{m-farhoudi@sbu.ac.ir}
 \affiliation{Department of Physics, Shahid Beheshti University, Evin, Tehran, 19839, Iran}

\begin{abstract}
\noindent
 While considering the chameleon scalar field model with the
spatially flat FLRW background, we investigate the late-time
acceleration phase of the universe, wherein we apply the typical
potential usually used in this model. Through setting some
constraints on the free parameters of the model, we indicate that
the non-minimal coupling between the matter and the scalar field
in such a model should be strongly coupled in order to have an
accelerated expansion of the universe at the late-time. We also
investigate the relative acceleration of the parallel geodesics by
obtaining the geodesic deviation equation in the context of
chameleon model. Then, through the null deviation vector fields,
we obtain the observer area-distance as a measurable quantity to
compare the model with other relevant models.
\end{abstract}
\date{7 July 2022}
\maketitle

\section{Introduction}
Nowadays, cosmology is facing with the most challenging problems
regarding accelerated expansion of the universe or dark energy as
well as dark mater. The mysterious acceleration of the universe
has been supported by the various cosmological observational
data~\cite{1}-\cite{parkinson}. As these observations are~not
consistent with the predictions of the Einstein gravity, this
theory has generally been amended/modified/generalized, in
particular to explain dark energy and dark matter (see, e.g.,
Refs.~\cite{nojiri0}-\cite{Bhattacharjee} and references therein).
For dark energy, numerous attempts have been performed, which
explain that the acceleration of the universe could have arisen
either from a dark energy component or being due to departure of
gravity from general relativity on cosmological scales. In the
former approach, dark energy models can be classified in two main
categories. The $\Lambda$CDM model (in which the universe contains
a constant energy density, cold dark matter and ordinary matter)
and the scalar field models with a dynamical equation of state,
see, e.g., Refs.~\cite{peebles}-\cite{bamba}. The first category
models have some difficulties, such as the cosmological constant
problem and the coincidence problem~\cite{carroll}-\cite{Bull}.
Some believe that the scalar field models are perhaps better
alternatives to the Einstein gravitational
theory~\cite{iess}-\cite{Faraoni}. In particular, quintessence is
a more general dynamical model in which the energy source of the
universe, unlike the cosmological constant, varies in space and
time~\cite{cald}-\cite{stein}. However, if one considers the
scalar field coupled to the matter in such theories, then a fifth
force and also large violation of the equivalence principle (EP)
will arise, whereas these results have~not been detected in the
solar system tests of gravity. To solve such a problem, the
chameleon model and its generalization have been
proposed~\cite{weltman1}-\cite{Quiros}.

In the chameleon model, as a scalar-tensor theory model for dark
energy, the scalar field couples minimally with geometry and
non-minimally with the matter field. It also uses an
environment-dependent screening mechanism that, although allows it
to deviate from general relativity at large scales, keeps it
consistent with the results of that at small scales. Indeed, the
scalar field couples to matter with a gravitational strength to
acquire a mass-dependent on the background matter density of the
environment, which leads its interaction to be effectively
short-ranged. Due to such a screening mechanism, the model remains
consistent with the tests of gravity on the terrestrial and the
solar system scales. The strength of force depends on the amount
of matter in the
environment~\cite{weltman1}-\cite{khoury}\cite{brax}-\cite{Burrage2017}.
In dense environments, such as on the earth, the force gets weaker
whose effects are barely detectable, and hence the theory will be
consistent with the experimental and observational tests in the
case of EP-violation and fifth force. On the other hand, as the
amount of matter decreases, the force becomes stronger. Hence, at
empty spaces, the force extends to a powerful range and one
expects detecting a fifth force and also the EP-violation.
However, despite some controversy, it is believed that the
chameleon field may play the role of dark energy causing the
cosmic late-time acceleration, see, e.g., Ref.~\cite{saba} and
references therein. In addition, the chameleon model during
inflation has also been investigated, see, e.g.,
Refs.~\cite{saba}-\cite{Sheikhahmadi}.

In the present work, we consider such a model and investigate the
late-time accelerated expansion of the universe. By applying the
typical potential usually used in the context of the chameleon
theory in the literature, we set some constraints on the free
parameters of the model such as the chameleon coupling constant
and the slope of the potential. Moreover, we generalize the
geodesic deviation equation (GDE) of the presented chameleon model
to probe the acceleration of the universe more instructive and
hopefully getting a better view of the cosmic acceleration phase.
Practically, the GDE offers a subtle understanding of the
structure of spacetime and characterizes the nature of
gravitational forces in an invariant procedure. The equation of
timelike geodesics in the Einstein frame receives a correction,
which is interpreted as the effects of a fifth force and the
violation of weak EP~\cite{fujji}. A physical definition of
geodesic is expressed as a trajectory of a body that is solely
under the influence of gravity, and mathematically, it is defined
as a curve that parallel transports its tangent vector. Actually,
the existence of a fifth force leads to modifications of the GDE.
Indeed, the GDE is one of the most significant equation in
gravitation that represents the effects of the curvature in the
spacetime and relates the Riemann curvature tensor to the relative
acceleration of two adjacent geodesics~\cite{Synge}-\cite{Ellis}.
It has also been studied in the contest of modified theories of
gravitation, see, e.g.,
Refs.~\cite{zare1,ShojaiShojai}-\cite{RasouliShojai} and
references therein.

The work is organized as follows. In the next section, we
introduce the chameleon model and obtain the field equations of
motion by taking the variation of the action. The cosmological
equations of the chameleon scalar field model are investigated in
Sec.~III, where matter-dominated phase and cosmic accelerated
phase are studied. In Sec.~IV, we investigate the GDE of this
model for the timelike and null geodesics within the spatially
flat Friedmann-Lema\^{\i}tre-Robertson-Walker (FLRW) background
and then, attain the corresponding Raychaudhuri equation and the
observer area-distance as a measurable physical quantity. At last,
we conclude the work in Sec.~V with the summary of the results.

\section{Chameleon Scalar Field Model}\label{Chameleon}
We start with the action of the chameleon scalar field model in
four dimensions as
\begin{equation}
S={S}_{\rm EH}+{S}_{\rm \phi}+{S}_{\rm \phi-m},
\end{equation}
where ${S}_{\rm EH}$ is the Einstein-Hilbert action, ${S}_{\phi}$
is the part of minimally coupled scalar field action and ${S}_{\rm
\phi-m}$ is the action representing coupling between the matter
field and the scalar field. More specifically, the action is
\begin{eqnarray}\label{action}
S\!\!\!&=\!\!\!&\!\!\int{\rm
d}^{4}x\sqrt{-g}\left(\frac{M^{2}_{\rm
Pl}R}{2}\right)\!-\!\!\!\int{\rm
d}^{4}x\sqrt{-g}\left[\frac{1}{2}\partial_{\mu}\phi\,
\partial^{\mu}\phi+V(\phi)\right]\nonumber
\\
 &&+\sum_{i}\int{\rm d}^{4}x\sqrt{-{\tilde g}^{(i)}}{L^{(i)}_{\rm m}}\left({\psi}^{(i)}_{\rm m},{\tilde g}^{(i)}_{\mu\nu}\right),
\end{eqnarray}
where $g$ is the determinant of the metric, $R$ is the Ricci
scalar, $M_{\rm Pl}\equiv(8\pi G)^{-1/2}\approx 10^{27}eV$ is the
reduced Planck mass (in the natural units, $\hbar=1=c$) and the
lower case Greek indices run from zero to three. Also, $V(\phi)$
is a self-interacting potential, $\psi^{(i)}$s are various matter
fields, ${L_{\rm m}^{(i)}}$s are Lagrangians of matter fields,
$\tilde g_{\mu\nu}^{(i)}$s are the matter field metrics that are
conformally related to the Einstein frame metric as
\begin{equation}
{\tilde g}^{(i)}_{\mu \nu}={e}^{ {2\frac {{\beta} _{i}\phi}{M_{\rm Pl}}}}{g}_{\mu \nu},
\end{equation}
where $\beta_{i}$s are dimensionless constants, which represent
different non-minimal couplings between the scalar field $ \phi $
and each matter species. However in this work, we just consider a
single matter component, and hence we omit the index $i$. The
scalar potential commonly used for the chameleon model in the
literature is the well-known run-away potential
\begin{equation}\label{v phi}
V\left( \phi  \right) = \frac{{{M^{4 + n}}}}{{{\phi ^n}}},
\end{equation}
with $M$ as some mass scale and $n$ as a positive or negative
integer constant.

The variation of action (\ref{action}) with respect to the metric
tensor $g_{\mu \nu}$ gives the field equations
\begin{equation}\label{G}
{G_{\mu \nu }} = \frac{1}{M_{\rm Pl}^2}\left(T_{\mu \nu
}^{(\phi)}+T_{\mu \nu }^{(m)}\right)= \frac{1}{M_{\rm
Pl}^2}\left(T_{\mu \nu }^{(\phi)}+ {e}^{ {2\frac
{{\beta}\phi}{M_{\rm Pl}}}}\tilde T_{\mu\nu}^{(m)}\right),
\end{equation}
where $ T_{\mu \nu }^{(\phi)} $ is the energy-momentum tensor of
the scalar field, namely
\begin{equation}\label{phiii}
T_{\mu \nu }^{(\phi)}  =  - \frac{1}{2}{g_{\mu \nu
}}{\partial^\alpha }\phi\, {\partial_\alpha }\phi  - {g_{\mu \nu
}}V\left( \phi  \right) + {\partial_\mu }\phi\, {\partial_\nu
}\phi,
\end{equation}
and $\tilde T_{\mu\nu}^{(m)}$ is the energy-momentum tensor of
matter in the Jordan frame that is conserved within this frame,
i.e. $\widetilde\nabla^{\mu} {\tilde T{^{(m)}_{\mu\nu}}}=0$, and
is defined as
\begin{equation}\label{tj}
\tilde T_{\mu\nu}^{(m)}=-\frac{2}{\sqrt{-\tilde
{g}}}\frac{\left(\delta{\sqrt{-\tilde {g}}}{L_{m}}\right)}{\delta
{\tilde g}^{\mu\nu}}.
\end{equation}
In addition, the variation of action (\ref{action}) with respect
to the scalar field gives
\begin{equation}\label{bax}
\Box\phi=\frac{{dV\left( \phi  \right)}}{{d\phi }}-\frac{{\beta}}{M_{\rm
Pl}}e^{4\frac{{\beta}\phi}{{M}_{\rm Pl}}}{\tilde
g}^{\mu\nu}{\tilde T}_{\mu\nu}^{(m)},
\end{equation}
where $\Box\equiv\nabla^{\alpha}\nabla_{\alpha}$ with respect to
the metric $g_{\mu\nu}$. Also, we assume the matter field as a
perfect fluid with the same state parameter $w^{(m)} $ in the both
frames. Hence, the trace of the energy-momentum tensor of matter
is
\begin{equation}
\tilde{T}^{(m)}={\tilde g}^{\mu\nu}\tilde
T_{\mu\nu}^{(m)}=-\left(1-3w^{(m)}\right)\tilde \rho^{(m)},
\end{equation}
and the relation of the matter density in the Einstein frame with
the one in the Jordan frame is
\begin{equation}\label{rhoo}
\rho^{(m)}= e^{4\frac{{\beta}\phi}{M_{\rm Pl}}}\tilde \rho^{(m)}.
\end{equation}
Note that $\left(1-3w^{(m)}\right)\rho^{(m)}=- {g^{\mu \nu
}}T_{\mu \nu }^{\left( m \right)} $, where $ T_{\mu \nu }^{\left(
m \right)}$ is the energy-momentum tensor of matter in the
Einstein frame that is~not covariantly conserved, i.e. $ {\nabla
^\mu }T_{\mu \nu }^{\left(m \right)} \ne 0 $. Moreover, for more
mathematical facilities, we can define a matter density as a
quantity independent of the chameleon scalar field $\phi$, which
is~not a physical matter density but is a conserved quantity
within the Einstein frame as~\cite{saba}
\begin{equation}\label{rh}
\rho(t)\equiv e^{3\left(1+w^{(m)}\right)\frac{{\beta}\phi}{M_{\rm
Pl}}}\tilde \rho^{(m)},
\end{equation}
and in turn obtain $ \rho^{(m)}=\rho\, e^{\left(1-3w^{(m)}\right)
\frac{\beta\phi}{M_{\rm Pl}}} $. Furthermore, substituting
relation~(\ref{rh}) into Eq.~(\ref{bax}) indicates that the scalar
field is dynamically governed by an effective potential, i.e.
\begin{equation}\label{field}
\Box\phi=\frac{{d{V_{\rm eff}}\left(\phi\right)}}{{d\phi }},
\end{equation}
with
\begin{equation}\label{eff}
{V}_{\rm eff}(\phi)\equiv V(\phi)+\rho\,
e^{\left(1-3w^{(m)}\right)\frac{\beta\phi}{M_{\rm
Pl}}}=V(\phi)+\rho^{(m)}
\end{equation}
that depends on the background matter density $\rho^{(m)}$ of the
environment.

\section{Cosmological Equations}\label{Cosmological}
In this section, we investigate the cosmological equations of the
chameleon scalar field model by considering the spatially flat
FLRW metric in the Einstein frame as
\begin{equation}\label{metric}
ds^{2}=-dt^{2}+a^{2}(t)\left(dx^{2}+dy^{2}+dz^{2}\right),
\end{equation}
where $t$ is the cosmic time, $a(t)$ is the scale factor
describing the cosmological expansion, and in turn
$\tilde{a}(t)=a(t)\exp({\beta\phi/M_{\rm Pl}})$ is the
corresponding one within the Jordan frame. Also, by accepting the
homogeneity and isotropy, and the scalar field being just a
function of the cosmic time, then by metric (\ref{metric}), the
field equation (\ref{field}) reads
\begin{equation}\label{phi}
\ddot{\phi}+3H\dot\phi+\frac{{d{V_{\rm eff}}\left(\phi\right)}}{{d\phi }} = 0,
\end{equation}
where $H(t)\equiv\dot{a}/a$ is the Hubble parameter and dot
denotes the derivative with respect to $t$. Moreover, by inserting
metric (\ref{metric}) into Eq.~(\ref{G}), it yields the
Friedmann-like equation as
\begin{equation}\label{friedmann}
{H^2} = \frac{1}{3{M_{\rm Pl}^2}}\left[ {\frac{1}{2}{{\dot \phi
}^2} + V\left( \phi \right) + \rho\, {e^{\left({1 - 3w^{(m)}}
\right)\frac{{\beta \phi }}{{{M_{\rm Pl}}}}}}} \right]
\end{equation}
and the generalized Raychaudhuri equation as
\begin{equation}\label{fried22}
\frac{{\ddot a}}{a}\! =\!  - \frac{1}{{3M_{\rm Pl}^2}}\! \left[\!
{{{\dot \phi }^2}\! -\! V\left( \phi  \right)\! +\! \left(\!
{\frac{{1\! +\! 3w^{(m)}}}{2}}\! \right)\!\rho {e^{\left( {1 -
3w^{(m)}} \right)\frac{{\beta \phi }}{{{M_{\rm Pl}}}}}}}\!
\right]\!.
\end{equation}

On the other hand, from relation (\ref{phiii}), one obtains the
energy density and the pressure density of the chameleon scalar
field as
\begin{equation}\label{rho phi 18}
{\rho ^{\left(\phi  \right)}} =
\frac{1}{2}{\dot \phi ^2} + V\left( \phi  \right)
\end{equation}
and
\begin{equation}\label{press phi 19}
{p^{\left(\phi  \right)}} = \frac{1}{2}{\dot \phi ^2}
- V\left( \phi  \right).
\end{equation}
Then, using Eqs. (\ref{rho phi 18}) and (\ref{press phi 19}),
Eqs.~(\ref{friedmann}) and (\ref{fried22}) read as
\begin{equation}\label{friedman}
{H^2} = \frac{1}{{3M_{\rm Pl}^2}}\left( {{\rho^{\left( \phi
\right)}} + {\rho ^{\left(m \right)}}} \right) =
\frac{1}{{3M_{\rm Pl}^2}} {\rho ^{\left( {\rm tot} \right)}}
\end{equation}
and
\begin{eqnarray}\label{fried2}
\frac{{\ddot a}}{a} &=&  - \frac{1}{{6M_{\rm Pl}^2}}\Big[ {{\rho
^{\left(\phi  \right)}} + {\rho ^{\left(m \right)}} + 3\left(
{{p^{\left(\phi  \right)}} + {p^{\left( m \right)}}} \right)}
\Big]\cr
 & =&  - \frac{1}{{6M_{\rm Pl}^2}}\left( {{\rho^{\left( {\rm
tot} \right)}} + 3{p^{\left({\rm tot} \right)}}} \right),
\end{eqnarray}
where $\rho^{(\rm tot)} \equiv \rho^{(\phi)}+\rho^{(m)}$ and
$p^{(\rm tot)} \equiv p^{(\phi)}+p^{(m)}$ are considered as the
total energy density and total pressure density, respectively.
Also, by employing Eqs. (\ref{phi}), (\ref{rho phi 18}) and
(\ref{press phi 19}), we obtain
\begin{equation}\label{rhophi}
\dot{\rho}^{(\phi)}+3H\left(\rho^{(\phi)}+p^{(\phi)}\right)=-X,
\end{equation}
and by the continuity equation for $\rho(t)$ in the Einstein
frame, i.e. $\dot{\rho}+3H\left(1+w^{(m)}\right)\rho=0$, we get
\begin{equation}\label{rhom}
\dot{\rho}^{(m)}+3H\left(\rho^{(m)}+p^{(m)}\right)=X,
\end{equation}
and in turn
\begin{equation}\label{rhotot}
\dot{\rho}^{(\rm tot)}+3H\left(\rho^{(\rm tot)}+p^{(\rm
tot)}\right)=0,
\end{equation}
where
\begin{equation}\label{x}
X\equiv\left(1-3w^{(m)}\right)\frac{\beta\dot{\phi}}{M_{\rm
Pl}}\rho e^{\left(1-3w^{(m)}\right)\frac{\beta\phi}{M_{\rm Pl}}}.
\end{equation}
The $X$ term acts as an interacting term among the scalar and
matter fields, which manifests itself as a deviation term into the
geodesic equation and is interpreted as some kind of internal
force among those. That is, due to the coupling between the scalar
and matter fields, the energy-momentum tensor of each one is~not
conserved. Nevertheless, the above relations indicate that,
although the energy density is~not separately conserved (and
conservation equations of the internal parts are~not independent),
its total is conserved as expected.

In the analysis of this work, we investigate the chameleon scalar
field during late-time of the universe. To proceed, we assume that
the evaluation of the chameleon scalar field with respect to time
being as the corresponding one considered in Ref.~\cite{saba1},
namely\footnote{In Ref.~\cite{saba1}, with their used scenario and
the specified function of the scalar field, the effects of
inflaton and chameleon have been described via one single scalar
field during the inflation and late-time.}
\begin{equation}\label{phiddotssaba}
\dot \phi  = \frac{{3(1+w^{(m)}){M_{\rm Pl}}}}{\beta
(1-3w^{(m)})}H.
\end{equation}
Also, we plausibly assume that the matter density is a
non-relativistic perfect fluid, i.e. dust matter with $w^{(m)}=0$,
and hence, relation (\ref{phiddotssaba}) reads
\begin{equation}\label{phi dot saba}
\dot \phi  = \frac{{3{M_{\rm Pl}}}}{\beta }H.
\end{equation}
Furthermore, we prefer to obtain the behavior of the chameleon
scalar field with respect to the redshift instead of time. For
this purpose, with the relation\footnote{The zero index indicates
the present time.}\
 $1+z=a_0/a$, we use the simple differential
operator
\begin{equation}\label{33}
\frac{d}{{dt}} = \frac{{da}}{{dt}}\frac{{dz}}{{da}}\frac{d}{{dz}}
=  - \left( {1 + z} \right)H\frac{d}{{dz}}.
\end{equation}
Thus, while employing relation (\ref{33}), relation (\ref{phi dot
saba}) gives
\begin{equation}\label{phiz}
\phi \left( z \right) = {\phi_0} - \frac{{3{M_{\rm Pl}}}}{\beta
}\ln \left( {1 + z} \right),
\end{equation}
where $ {\phi_0} $ is an integration constant that is equal to $
\phi \left( z \right) $ for $z = 0$.

Now, to obtain the total (or, the effective) state parameter of
this chameleon model, we start by the dimensionless density
parameters defined as
\begin{equation}\label{om0}
{\Omega_{\dot{\phi}}}\equiv\frac{{{{\dot \phi }^2}}}{2{\rho
_0^{\rm (crit)}}}, \quad\ \ {\Omega_{V}} \equiv
\frac{V(\phi)}{{\rho _0^{\rm (crit)}}} \quad\ {\rm and} \quad\
{\Omega_m} \equiv \frac{\rho^{(m)}}{{\rho _0^{\rm (crit)}}},
\end{equation}
where $ \rho _0^{\rm (crit)} \equiv 3H_0^2M_{\rm Pl}^2 $ is the
critical density of the universe at the present time. Hence, in
general case, Eq. (\ref{friedman}) can be rewritten as
\begin{equation}\label{H2}
{H^2} = H_0^2\left( {{\Omega_{\dot{\phi}}}+{\Omega_V }+{\Omega_m}} \right),
\end{equation}
and the total state parameter of this model is
\begin{equation}\label{wt}
{w^{\left({\rm tot} \right)}} \equiv \frac{{{p^{\left( {\rm tot}
\right)}}}}{{{\rho ^{\left({\rm tot} \right)}}}} = \frac{{{\Omega
_{\dot \phi }} - {\Omega _V}}}{{{\Omega _{\dot \phi }} + {\Omega
_V} + {\Omega _m}}},
\end{equation}
and equivalently
\begin{equation}\label{state}
{w^{\left({\rm tot} \right)}} = \frac{{H_0^2}}{{{H^2}}} \left(
{{\Omega_{\dot \phi }} - {\Omega_V}} \right).
\end{equation}
Substituting relation (\ref{phi dot saba}), for the case of
$w^{(m)}=0$, into the first definition (\ref{om0}) leads to
\begin{equation}\label{omega dot}
{\Omega _{\dot \phi }} =\frac{3{H^2}}{2{\beta ^2}H_0^2}
\end{equation}
and then, inserting it into relation (\ref{state}) yields
\begin{equation}\label{state T}
{w^{\left( {\rm tot} \right)}} = \frac{3}{{2{\beta ^2}}} -
\frac{{H_0^2}}{{{H^2}}}{\Omega _V}.
\end{equation}

At this stage, we manage to get $ {\Omega _V} $ and $ H^{2} $ in
terms of $ z $. For this purpose, inserting function~(\ref{phiz})
into (\ref{v phi}) gives
\begin{equation}\label{omega v}
{\Omega_V} = {\Omega_{\left( 0 \right)V}}{\left(
{\frac{{{\phi_0}}}{\phi }} \right)^n} = {\Omega_{\left( 0
\right)V}}{\left[ {1 - \frac{{3{M_{\rm Pl}}}}{{\beta {\phi_0}}}\ln
\left( {1 + z} \right)} \right]^{ - n}},
\end{equation}
where
\begin{equation}
{\Omega _{\left( 0 \right)V}} = \frac{{V\left( {{\phi_0}}
\right)}}{{\rho_0^{\rm (crit)}}} = \frac{1}{{\rho_0^{\rm
(crit)}}}\left( {\frac{{{M^{4 + n}}}}{{\phi_0^n}}} \right).
\end{equation}
Also, for the conserved matter within the Jordan
frame\footnote{Its related continuity equation is
$\dot{\tilde\rho}^{(m)}+3\tilde H\left(\tilde\rho^{(m)}+\tilde
p^{(m)}\right)=0$, where $\tilde H=H+3\beta\dot{\phi}/M_{\rm
Pl}$.}\
 with $w^{(m)}=0$, we have ${\tilde \rho ^{\left(m \right)}} =
\tilde \rho _0^{\left( m \right)} {\left( {1 + z} \right)^3} $,
hence within the Einstein frame, considering function (\ref{phiz})
while using (\ref{rhoo}), we obtain
\begin{equation}\label{rho mz}
{\Omega _m} = {\tilde \Omega _{\left( 0 \right)m}} \,
{e^{4\frac{{\beta {\phi _0}}}{{{M_{\rm Pl}}}}}}{\left( {1 + z} \right)^{ - 9}},
\end{equation}
where $ {\tilde \Omega _{\left( 0 \right)m}}\equiv \tilde \rho
_0^{\left(m \right)}/\rho _0^{\rm (crit)} $. Then, using
relations~(\ref{omega dot}), (\ref{omega v}) and (\ref{rho mz}),
Eq.~(\ref{H2}) reads
\begin{eqnarray}\label{H2 z}
H^2 = \frac{H_0^2}{1 - \frac{3}{2\beta^2}}
 \Bigg\{ &{}&\!\!\!\!\!\!\!\!\!\!\!\Omega_{(0)V}\left[ 1 -
 \frac{3M_{\rm Pl}}{\beta \phi_0}\ln (1 + z) \right]^{- n}\cr
 \!\!\!\!\!& +&\! \tilde \Omega_{( 0 )m} \, e^{4\frac{\beta
 \phi_0}{M_{\rm Pl}}}( 1 + z)^{-9}\Bigg\},
\end{eqnarray}
for when $\beta\neq \pm\sqrt{3/2}$. Finally, substituting relation
(\ref{omega v}) and Eq. (\ref{H2 z}) into relation (\ref{state T})
gives the total state parameter for non-relativistic perfect
fluids, in the case of dust matters, in terms of the redshift as
\begin{eqnarray}\label{w z}
{w^{\left({\rm tot} \right)}} =&{}&\!\!\! \frac{3}{{2{\beta ^2}}}
- \left( {1 - \frac{3}{{2{\beta^2}}}} \right)\Bigg\{ 1 +
\frac{{{\tilde \Omega_{\left( 0 \right)m}} \, {e^{4\beta
{\phi_0}/{M_{\rm Pl}}}} }}{{{\Omega_{\left( 0\right)V}}
  {\left( {1 + z} \right)^{ 9}} }}\times\cr
&{}&\qquad {{\left[ {1 - \frac{{3{M_{\rm Pl}}}}{{\beta
{\phi_0}}}\ln \left( {1 + z} \right)} \right]}^n}\Bigg\}^{-1}.
\end{eqnarray}

Meanwhile, let us also employ the deceleration parameter, which is
a dimensionless measure of the cosmic acceleration of the
expansion of the universe, defined as $ q \equiv  - \ddot aa/{\dot
a^2} = - \ddot a/(a{H^2}) $. In this regard, inserting Eqs.
(\ref{friedman}) and (\ref{fried2}) into the definition of $ q $
and using the definition of the total state parameter, leads to
\begin{equation}\label{decel}
q = \frac{{1 + 3{w^{\left({\rm tot} \right)}}}}{2}.
\end{equation}
Obviously, if $ q>0 $, then the expansion of the universe will be
a decelerated one, and if $ q<0 $, then it will be an accelerated
one. The transition point from the deceleration to the
acceleration phase is when $ q=0 $, and in this case, relation
(\ref{decel}) shows that it corresponds to $w_{\rm
trans.}^{\left({\rm tot} \right)}=-1/3 $, as expected. Hence, in
general case, using relation (\ref{state}), it obviously gives
$\left[\Omega_{\dot \phi }- \Omega_V=-H^2/(3H_0^2)\right]_{\rm
trans.}$, and in turn, $\left[p^{(\phi)}=-H^2M^2_{\rm
Pl}\right]_{\rm trans.}$. Also, for the case of dust matters with
$w^{(m)}=0$, using relation (\ref{state T}), it yields
$\left[V=H^2M^2_{\rm Pl}(9+2\beta^2)/(2\beta^2)\right]_{\rm
trans.}$. We pursue this issue to discuss about the transition
redshift in the next section.

We have plotted relation (\ref{w z}) in Figure~$1$ to obtain the
allowed values of the $n$ parameter in this model for $ \beta=3.7
\times {10^2} $ and $ \beta=1 $ with initial value $ {\phi _0} =
{\beta ^{ - 1}} $ (regardless of units) as assumptions. Note that,
the value $ \beta=3.7 \times {10^2} $ is an upper bound on this
parameter in the chameleon model that is consistent with the
experimental constrain obtained in Ref.~\cite{Jaffe}.
%%%%%%%%%%%%%%%%%%%%%%%%%%%%%%
\begin{figure}[h]
\includegraphics[scale=0.9]{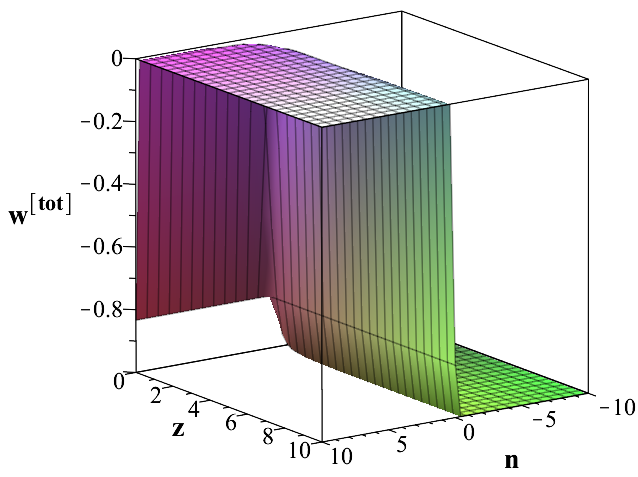}
\includegraphics[scale=0.9]{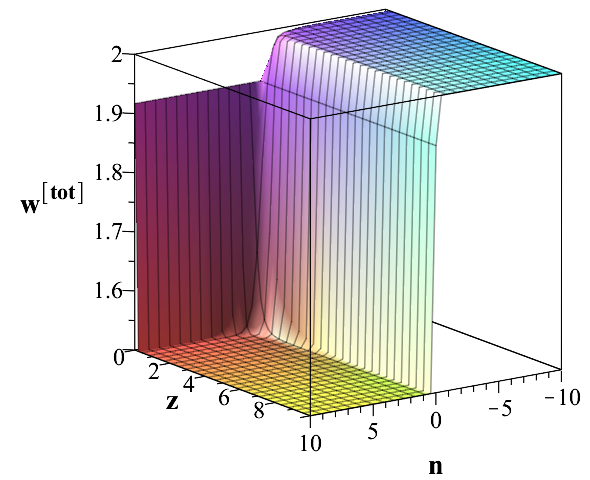}
\caption{The plot of the total state parameter with assumptions $
{ \tilde \Omega }_{\left( 0 \right)m}/\Omega_{\left( 0 \right)V}
\simeq3/7$, $ {M_{\rm Pl}}=1 $ and, regardless of units, ${\phi
_0} = {\beta ^{ - 1}} $. Besides, the top and bottom diagrams have
been plotted with $ \beta =3.7 \times {10^2} $ and $ \beta = 1 $,
respectively.}
\end{figure}
The top diagram in Figure~$1$ indicates that, for the case $ \beta
=3.7 \times{10^2} $ with $ n>0 $, the total state parameter in the
present time, i.e. at $ z=0 $, starts to increase from the value
less than $ -1/3 $ with increasing redshift, which is a suitable
model for the evolution of the universe. In addition, the presence
of constraint $n>0$ is consistent with the inverse power-law
potentials first considered in the original suggestion for the
chameleon model~\cite{weltman1}. Whereas, the bottom diagram in
this figure, for $-10<n<10$ and $0<z<10$, shows unacceptable
results for the case $ \beta = 1 $ because, at the present time,
one expects the total state parameter to be less than $ -1/3$.
However, with $ n<0 $, the top diagram in Figure~$1$ also
illustrates not a true model because by increasing redshift from $
z=0 $ up to $ z=10 $, the total state parameter does~not increase.

\subsection{Matter-Dominated Phase}
In the matter-dominated phase of the universe, i.e. under
assumption ${\rho^{\left(m \right)}} \gg {\rho^{\left(\phi
\right)}} $ that corresponds to $ {\Omega _m} \gg {\Omega _{\dot
\phi }}+{\Omega _V} $, Eq.~(\ref{H2}) leads to
\begin{equation}
{H^2} \simeq H_0^2 \Omega {_m}
\end{equation}
and in turn, the state parameter from relation~(\ref{state}) gives
\begin{equation}\label{w matter}
{w^{\left({\rm tot} \right)}} \simeq \frac{{{\Omega _{\dot \phi
}} - {\Omega _V}}}{{{\Omega _m}}}.
\end{equation}
In situations that still\footnote{From (\ref{phiz}), while
regardless of units assuming ${\phi _0} = {\beta^{ - 1}} $, when
the scalar $\phi$ has positive values, from (\ref{v phi}) will
also be the potential $V$. Hence, from $ {\Omega _m} \gg {\Omega
_{\dot \phi }}+{\Omega _V} $, one obviously has $ {\Omega _m} \gg
{\Omega_{\dot \phi }}-{\Omega _V} $.}\
 $ {\Omega _m} \gg {\Omega_{\dot \phi }}-{\Omega _V} $, then $w^{\left({\rm tot}
\right)}\simeq 0$, which is consistent with the expected value in
the matter-dominated epoch of the universe.

\subsection{Cosmic Accelerated Phase}
Since the dust matter density decreases over the time, one can
plausibly assume the chameleon scalar field-dominated phase, i.e.
$ {\rho ^{\left(\phi \right)}} \gg {\rho ^{\left( m \right)}} $,
at the late-time universe. Under such an assumption,
Eq.~(\ref{H2}) leads to
\begin{equation}\label{H2 acc}
{H^2} \simeq H_0^2\left( {{\Omega _{\dot \phi }}
+ {\Omega _V}} \right).
\end{equation}
Then, inserting Eq.~(\ref{H2 acc}) into relation (\ref{state T})
gives the state parameter at the late-time accelerating phase to
be
\begin{equation}\label{w acc}
{w^{\left({\rm tot} \right)}} \simeq \frac{3}{{2{\beta ^2}}} -
\frac{{{\Omega _V}}}{{{\Omega _{\dot \phi }} + {\Omega _V}}}.
\end{equation}
In this phase of the universe, if we consider $ \beta=1 $ while
inserting Eq.~(\ref{H2 acc}) into relation~(\ref{omega dot}) that
yields $ {\Omega_{\dot \phi }} = -3{\Omega_V} $, then
relation~(\ref{w acc}) will give $ {w^{\left({\rm tot}
\right)}}=2 $ that is inconsistent with the observations at the
late-time universe. Whereas with the value $ \beta =3.7 \times
{10^2} $, relation~(\ref{omega dot}), while inserting Eq.~(\ref{H2
acc}) into it, renders $ {\Omega _{\dot \phi }} \simeq {10^{ -
5}}\,\Omega_V $, and in turn in this case for dust matters,
relation (\ref{w acc}) gives
\begin{equation}
{w^{\left({\rm tot} \right)}} \simeq \frac{3}{{2{\beta ^2}}} - 1
\simeq {10^{ - 5}} - 1 \simeq  - 1.
\end{equation}

Accordingly, the analysis shows that in order to have a viable
chameleon model with $w^{(m)}=0$ during the acceleration phase of
the universe at the late-time universe, the coupling constant
between the matter and scalar fields should be much greater than
unity. In this respect, it is worth mentioning that, although
Weltman and Khoury, in their original suggestion for the chameleon
model, considered the possibility of coupling the scalar field to
the matter field with the gravitational strength of the order of
unity, Mota and Shaw showed that the scalar field theories with
strongly coupling are viable due to the non-linearity effects of
the theory~\cite{mota1,mota2}.

\section{Geodesic Deviation Equation}\label{GDE}
The Einstein field equations specify how the curvature depends on
the matter sources, where one can obtain the consequences of the
spacetime curvature through the GDE. Formulating the cosmological
equation in the GDE form would be a model independent way. Hence,
in this section, in order to probe the acceleration of the
universe at the late-time more instructive, we derive the GDE in
the presented chameleon model. For this purpose, we start from the
general expression for the GDE~\cite{Hawking, Faraoni82}
\begin{equation}\label{eta1}
\frac{{{D^2}{\eta^\mu }}}{{D{\nu^2}}} = R^\mu{}_{\nu \alpha \beta
} {V^\nu }{V^\alpha }{\eta^\beta }
 - \frac{\beta }{{{M_{\rm Pl}}}}{\eta ^\alpha }{\nabla _\alpha }\left( {{\partial^\mu }\phi } \right),
\end{equation}
with the parametric $x^\mu(s,\nu)$, where $s$ labels distinct
geodesics, the parameter $ \nu $ is an affine parameter along the
geodesic and ${\eta ^\mu } = \partial x^\mu/\partial s $ is the
orthogonal deviation vector of two adjacent geodesics. Also, $
D/D\nu $ is the covariant derivative along the curve and the
normalized vector field $ {V^\mu } = \partial{x^\mu }/\partial\nu
$ is tangent to the geodesics. The second term, which appears in
the right side of this equation, illustrates that, in general,
there is a fifth force mediated by $ \phi $, which acts on any
massive test particle. However, as we have assumed that the
universe is isotropic and homogeneous, only the time derivatives
of the scalar field do~not vanish, and also in the comoving frame,
one has $ \eta^{0}=0 $, hence in this case, Eq.~(\ref{eta1})
reads~\cite{wald,RasouliShojai}
\begin{equation}\label{eta}
\frac{{{D^2}{\eta ^\mu }}}{{D{\nu ^2}}} = R^\mu{}_{\nu \alpha
\beta } {V^\nu }{V^\alpha }{\eta ^\beta }.
\end{equation}
Now, to attain the relation
between the geometrical properties of the spacetime with the field
equations governed from the chameleon model, we use the expression
of the Riemann tensor in 4-dimensions, namely
\begin{eqnarray}\label{Riman}
{R_{\mu \nu \alpha \beta }} = &{}&\!\!\!\!{C_{\mu \nu \alpha \beta
}} + \frac{1}{2} \Big( {g_{\mu \alpha }}{R_{\nu \beta }} - {g_{\mu
\beta }}{R_{\nu \alpha }} + {g_{\nu \beta }}{R_{\mu \alpha }}\cr
&{}&\!\!\!\! -{g_{\nu \alpha }}{R_{\mu \beta }} \Big) -
\frac{1}{6}R\Big( {{g_{\mu \alpha }}{g_{\nu \beta }} - {g_{\mu
\beta }}{g_{\nu \alpha }}} \Big),
\end{eqnarray}
and note that, in the case of the FLRW metric, the corresponding
Weyl tensor $ {C_{\mu \nu \alpha \beta }}  $ is zero. In addition,
for $w^{(m)}=0$ with $ T_{\mu \nu }^{(m)} = {\rho ^{(m)}}{u_\mu
}{u_\nu }=\left(\rho\, {e^{\beta \phi /{M_{\rm Pl}}}}\right)u_\mu
{u_\nu } $, where $ {u^\mu } $ is the comoving unit velocity
vector to the matter flow, one easily obtains the Ricci tensor
from Eq. (\ref{G}) as
\begin{eqnarray}
{R_{\mu \nu }} = \frac{1}{{M_{\rm Pl}^2}}\Big[
&{}&\!\!\!\!\frac{1}{2}{g_{\mu \nu }}{{\dot \phi }^2}  - {g_{\mu
\nu }} V(\phi) + {\partial_\mu }\phi\, {\partial_\nu }\phi \cr
 &{}&\!\!\!\! + \rho\, {e^{^{\beta
\phi /{M_{\rm Pl}}}}}{u_\mu }{u_\nu }\Big] + \frac{1}{2}{g_{\mu
\nu}}R,
\end{eqnarray}
and in turn with $ u_{\mu} u^{\mu}=-1 $, the Ricci scalar as
\begin{equation}\label{nas1}
R = \frac{1}{{M_{\rm Pl}^2}}\left[ { - {{\dot \phi }^2} + 4V(\phi
) + \rho\, {e^{\beta \phi /{M_{\rm Pl}}}}} \right].
\end{equation}
Inserting these relations into relation (\ref{Riman}) gives
\begin{eqnarray}
{R_{\mu \nu \alpha \beta }}=&\!\!\!\!\!\!\frac{1}{{2M_{\rm Pl}^2}}
\Big( {g_{\mu \alpha }}{\partial_\nu }\phi\, {\partial_\beta }\phi
 - {g_{\mu \beta }}{\partial_\nu }\phi\, {\partial_\alpha }\phi
 \cr
&\qquad\ + {g_{\nu \beta }} {\partial_\mu }\phi\, {\partial_\alpha
}\phi - {g_{\nu \alpha }}{\partial_\mu }\phi\, {\partial_\beta
}\phi \Big)\cr
 &\!\!\!\! + \frac{{ \rho\,
{e^{\beta \phi /{M_{\rm Pl}}}}}}{{2M_{\rm Pl}^2}} \Big( {g_{\mu
\alpha }}{u_\nu }{u_\beta } - {g_{\mu \beta }}{u_\nu }{u_\alpha
}\cr
 &\qquad\qquad\  + {g_{\nu \beta }}{u_\mu
}{u_\alpha } - {g_{\nu \alpha }}{u_\mu }{u_\beta } \Big)\cr
 &\!\!\!\! + \frac{1}{{3M_{\rm Pl}^2}}\Big[ {\frac{1}{2}{{\dot
\phi }^2} + V \left( \phi  \right) + \rho\, {e^{\beta \phi
/{M_{\rm Pl}}}}}\Big]\times \cr
 &\Big(
{{g_{\mu \alpha }}{g_{\nu \beta }} - {g_{\mu \beta }}{g_{\nu
\alpha }}} \Big),
\end{eqnarray}
and then, under conditions $ {\eta ^0 }=0 $ in the comoving frame
and ${\eta _\mu }{u^\mu } = 0 = {\eta _\mu}{V^\mu} $, we obtain
\begin{eqnarray}\label{R}
\!\!\!\!\!\!\!\!\!R{^\mu}_{\nu \alpha \beta }&&\!\!\!\!\!\! {V^\nu
}{V^\alpha }{\eta ^\beta} =\cr
 &&\!\!\!\!\!\!\! -\frac{1}{{2M_{\rm Pl}^2}}
\Big[ {{\left( {{V^\nu }{\partial_\nu }\phi } \right)}^2} + \rho\,
{e^{\beta \phi /{M_{\rm Pl}}}}{{\left( {{u_\nu }{V^\nu }}
\right)}^2}\Big]{\eta^\mu }\cr
 &&\!\!\!\!\!\!\! -
\frac{1}{3M_{\rm Pl}^2} \Big[ {\frac{1}{2}{{\dot \phi }^2} + V
\left( \phi  \right) + \rho\, {e^{\beta \phi /{M_{\rm Pl}}}}}
\Big] {V^\nu }{V_\nu } {\eta ^\mu }.
\end{eqnarray}
Accordingly, by employing the total energy $ E=-u^{\mu}V_{\mu} $,
$\varepsilon \equiv {V^\mu }{V_\mu } $, relation (\ref{nas1}) and
the definitions $\rho^{(\rm tot)} $ and $ p^{(\rm tot)} $ for
$w^{(m)}=0$ into relation (\ref{R}), Eq.~(\ref{eta}) reads
\begin{eqnarray}\label{gde}
\frac{{{D^2}{\eta ^\mu }}}{{D{\nu ^2}}}&&\!\!\!\!\! +
\frac{1}{{2M_{\rm Pl}^2}} \Big[{\left( {\dot \phi {V^0}}
\right)^2} + {\rho ^{\left( m \right)}}{E^2}\qquad\qquad\cr
 &&\!\!\!\!\! +
\frac{\varepsilon}{3}\Big( {{\rho ^{\left( {\rm tot} \right)}}
 + 3{p^{\left( {\rm tot} \right)}} + M_{\rm Pl}^2R} \Big) \Big]{\eta ^\mu } = 0.
\end{eqnarray}
The four-velocity of FLRW comoving observers is $ u^\mu=(1,0,0,0)
$, hence $ E=V^0 $, and finally the GDE for the presented
chameleon model for dust matters with $w^{(m)}=0$ is
\begin{eqnarray}\label{gdee}
\frac{{{D^2}{\eta ^\mu }}}{{D{\nu ^2}}}&&\!\!\!\!\! +
\frac{1}{{2M_{\rm Pl}^2}}\Big[ {E^2} \left( {{\rho ^{\left({\rm
tot} \right)}} + {p^{\left({\rm tot} \right)}}} \right) \qquad \cr
 &&\!\!\!\!\! + \frac{\varepsilon }{3}\left( {{\rho ^{\left({\rm tot}
\right)}} + 3{p^{\left( {\rm tot} \right)}} + M_{\rm Pl}^2R}
\right) \Big]{\eta ^\mu } = 0.
\end{eqnarray}

\subsection{GDE for Timelike Vector Fields}
For timelike vector fields corresponded to the comoving observers
within the FLRW background, the affine parameter $ \nu $ is
actually the proper time $ t $. Hence in this case, $
\varepsilon=-1 $, $ E=1 $  and GDE~(\ref{gdee}) reads
\begin{equation}
\frac{{{D^2}{\eta ^\mu }}}{{D{t^2}}} + \frac{1}{{6M_{\rm
Pl}^2}}\left( {2{\rho ^{\left({\rm tot} \right)}} - M_{\rm
Pl}^2R} \right){\eta ^\mu } = 0,
\end{equation}
which in turn, by substituting $\rho ^{\left( {\rm tot} \right)}$
and relation (\ref{nas1}), in the case of dust matters, we obtain
\begin{equation}\label{free}
\frac{{{D^2}{\eta ^\mu }}}{{D{t ^2}}} + \frac{1}{{6M_{\rm
Pl}^2}}\left( {2{{\dot \phi }^2} - 2V\left( \phi  \right) + \rho\,
{e^{\beta \phi /{M_{\rm Pl}}}}} \right){\eta ^\mu } = 0.
\end{equation}
On the other hand, with respect to the comoving tetrad frame, the
deviation vector can be rewritten as $ {\eta ^\mu } = a\left( t
\right){e^\mu } $, and in addition, due to the isotropy of the
spacetime, one has
\begin{equation}
\frac{{D{e^\mu }}}{{Dt}} = 0.
\end{equation}
Therefore, Eq. (\ref{free}) becomes
\begin{equation}\label{ddot}
\frac{{\ddot a}}{a} =  - \frac{1}{{6M_{\rm Pl}^2}}\left( {2{{\dot
\phi }^2} - 2V\left( \phi \right) + \rho\, {e^{\beta \phi /{M_{\rm
Pl}}}}} \right),
\end{equation}
which is a particular case of the Raychaudhuri dynamical
Eq.~(\ref{fried22}) for $w^{(m)}=0$, as expected to be true even
for any value of it. Clearly, Eq.~(\ref{free}) illustrates that in
order to have an accelerated expansion of the universe, two
adjacent geodesics should recede from each other, i.e. $ {D^2
\eta^\mu} / Dt^2 >0 $, that corresponds to ${\ddot a}>0  $ in
Eq.~(\ref{ddot}).

In the following, we proceed to obtain the value of the transition
redshift from the deceleration to the acceleration phase of the
universe in the chameleon model for dust matters (i.e.,
$w^{(m)}=0$) with a strong coupling. For this purpose, as the
transition point is when $ {\ddot a}=0 $, thus according to
Eq.~(\ref{ddot}), it occurs when
\begin{equation}\label{trans}
\left[{\Omega _{\dot \phi}} = \frac{\Omega _V}{2} - \frac{\Omega
_m}{4}\right]_{\rm trans.}.
\end{equation}
By inserting relation (\ref{trans}) into Eq. (\ref{H2}), it gives
\begin{equation}\label{H2zz}
\left[{H^2} = \frac{3}{2}H_0^2\left( {{\Omega _V} + \frac{{{\Omega
_m}}}{2}} \right)\right]_{\rm trans.},
\end{equation}
and in turn, relation (\ref{state T}) reads
\begin{equation}\label{wtz}
\left[{w^{\left({\rm tot} \right)}} = \frac{3}{{2{\beta ^2}}} -
\frac{4}{{6 + 3\Omega_m/\Omega_V}}\right]_{\rm trans.}.
\end{equation}
On the other hand, as at the transition point $w_{\rm
trans.}^{\left({\rm tot} \right)}=-1/3$, hence with the strong
coupling $ \beta =3.7 \times{10^2} $, one approximately
obtains\footnote{Indeed, its value is less than $2$, i.e.
$2-\epsilon$, which in turn by relation (\ref{trans}) gives
$\left[{\dot \phi}^2=\epsilon V/2\right]_{\rm trans.}$, where $V$
at the transition point is a positive value as given under
relation (\ref{decel}).}\
 $\left[\Omega_m/\Omega_V\right]_{\rm trans.}\simeq 2$. By
substituting relations (\ref{omega v}) and (\ref{rho mz}) into
this result, we achieve the constraint
\begin{equation}\label{constraintz}
\frac{{\tilde \Omega }_{\left( 0 \right)m} \, e^{4\beta
\phi_0/M_{\rm Pl}}}{{{\Omega _{\left( 0 \right)V}} \, {{\left( {1
+ z_{\rm trans.}} \right)}^9}}} {\left[ {1 - \frac{{3{M_{\rm
Pl}}}}{{\beta {\phi _0}}}\ln \left( {1 + z_{\rm trans.}} \right)}
\right]^n}\simeq 2.
\end{equation}
Now, with assumptions $ \beta =3.7 \times{10^2} $, ${ \tilde
\Omega }_{\left( 0 \right)m}/\Omega_{\left( 0 \right)V} \simeq
3/7$, ${M_{\rm Pl}}=1 $ and (regardless of units) ${\phi _0} =
{\beta ^{ - 1}} $, we can attain the transition redshift in the
chameleon model. Indeed, Eq. (\ref{constraintz}) gives the value
of the transition redshift\footnote{For negative values of the $n$
parameter, the value of the transition redshift is an imaginary
number.}\
 to be $ z_{\rm trans.} \simeq 0.202$ with $ n=1 $, $ z_{\rm
trans.} \simeq 0.157$ with $ n=2 $, $ z_{\rm trans.} \simeq 0.097$
with $ n=5 $, and $ z_{\rm trans.} \simeq 0.060$ with $ n=10 $.
These obtained values are less than the corresponding value of the
$ \Lambda \rm CDM $ model\rlap.\footnote{In the $ \Lambda \rm CDM
$ model, the calculation of the transition redshift yields $
z_{\rm trans.} \simeq 0.67$.}\
 Also, in Ref.~\cite{zare1}, the
corresponding value obtained in the $f(R,T) $ gravity model is $
z_{\rm trans.} =0.11 $. Hence, the obtained results indicate that
the coupling between the matter field and the scalar field in the
chameleon model (and also between the matter field and geometry in
the $ f(R,T) $ gravity model) leads to have the matter-dominated
phase ending later than a model without considering such coupling.

It is worth noting that the chameleon theory, due to the
non-minimal coupling between the matter and scalar field, has led
to an additional force. Indeed, the equation of timelike geodesics
shows such a fifth force by receiving an amendment in the Einstein
frame. In other words, the GDE shows that, although a freely
falling particle appears to be at rest in a comoving frame falling
with the particle, a pair of nearby freely falling particles
indicates a relative motion that can reveal the presence of a
gravitational field to an observer that falls with those.

\subsection{GDE for Null Vector Fields}
In this subsection, we investigate the past-directed null vector
fields. With the FLRW metric, for null vector fields, one has $
\varepsilon  = 0 $. Also, by considering ${\eta ^\mu } = \eta\,
{e^\mu } $ and using a parallelly propagated and aligned
coordinate basis, we have $ D{e^\mu }/D\nu  = {V^\alpha
}{\nabla_\alpha }{e^\mu } = 0 $, and hence GDE (\ref{gdee}) (for
dust matters) reduces to
\begin{equation}\label{nul geo}
\frac{{{d^2}\eta }}{{d{\nu ^2}}} + \frac{1}{{2M_{\rm Pl}^2}}\left(
{{\rho ^{\left( {\rm tot} \right)}} + {p^{\left({\rm tot}
\right)}}} \right){E^2}\eta  = 0,
\end{equation}
and actually, using definitions $\rho^{({\rm tot})}$ and $p^{({\rm
tot})}$ for $w^{(m)}=0$, it reads
\begin{equation}
\frac{{{d^2}\eta }}{{d{\nu ^2}}} + \frac{1}{{2M_{\rm Pl}^2}}\left(
{{{\dot \phi }^2} + \rho\, {e^{\frac{{\beta \phi }}{{{M_{\rm
Pl}}}}}}} \right){E^2}\eta  = 0.
\end{equation}
Regarding the focusing condition mentioned in
Refs.~\cite{Schneider,Ellis}, all the past-directed null geodesics
from Eq.~(\ref{nul geo}) will experience focusing if the condition
$ {\rho ^{\left( {\rm tot} \right)}}+{p ^{\left({\rm tot}
\right)}}>0 $ is met. For the chameleon model in this case, this
condition is actually
\begin{equation}\label{focus}
{{{\dot \phi }^2} + \rho {e^{\frac{{\beta \phi }}{{{M_{\rm
Pl}}}}}}}>0,
\end{equation}
which, as the density of matter is always positive, is confirmed.

In continuation, it is more suitable to write the GDE for null
vector fields as a function of the redshift to obtain the observer
area-distance in the chameleon model. To perform this task, we
start from the differential operator
\begin{equation}
\frac{d}{{d\nu }} = \frac{{dz}}{{d\nu }}\frac{d}{{dz}}
\end{equation}
that obviously yields
\begin{eqnarray}\label{eq63}
\frac{d^2}{d\nu^2}\!\!&=&\!\!\left(\frac{dz}{d\nu}\right)^2\frac{d^2}{dz^2}+
\frac{d^2z}{d\nu^2}\frac{d}{dz}\cr
 \!\!&
=&\!\!\left(\frac{d\nu}{dz}\right)^{-2}\left[ \frac{d^2}{dz^2} -
(\frac{d\nu}{dz})^{- 1}\frac{d^2\nu}{dz^2}\frac{d}{dz}\right].
\end{eqnarray}
Then, using the relation $(1+z) = a_0/a = E/E_0$ with assumption $
a_0=1 $ for the present-time, one obtains~\cite{Ellis}
\begin{equation}\label{eq66}
dz =  - \left( {1 + z}\right)\frac{{\dot a}}{a}\frac{{dt}}{{d\nu
}}d\nu  =  - \left( {1 + z} \right)HEd\nu,
\end{equation}
that can be rearranged into
\begin{equation}\label{eq67}
\frac{{d\nu }}{{dz}} = - \frac{1}{{{E_0}H{{\left( {1 + z}
\right)}^2}}},
\end{equation}
and in turn, get
\begin{equation}\label{eq68}
\frac{{{d^2}\nu }}{{d{z^2}}} = \frac{1}{{{E_0}H{{\left({1 + z}
\right)}^3}}}\left[ 2 + \frac{{\left( {1 + z} \right)}}{H}\left(
{\frac{dH}{dz}} \right) \right].
\end{equation}
By inserting the relation
\begin{equation}\label{eq69}
\frac{dH}{dz} = \frac{d\nu }{dz}\frac{{dt}}{d\nu }\frac{dH}{dt}=
- \frac{\dot H}{H\left( {1 + z}\right)}
\end{equation}
into relation~(\ref{eq68}), it reads
\begin{equation}\label{eq70}
\frac{{{d^2}\nu }}{{d{z^2}}} = \frac{1}{{{E_0}H{{\left( {1 +
z}\right)}^3}}}\left( {2 - \frac{{\dot H}}{{{H^2}}}} \right).
\end{equation}
By substituting $\dot H ={\ddot a}/a - {H^2}$ into
relation~(\ref{eq70}) while using Eq.~(\ref{fried2}), one obtains
\begin{equation}\label{eq72}
\frac{{{d^2}\nu }}{{d{z^2}}} = \frac{1}{{{E_0}H{{\left( {1 +
z}\right)}^3}}}\left( {3 + \frac{{
 {{\rho ^{\left( {\rm tot} \right)}} + 3{p^{\left(
{\rm tot} \right)}}}}}{{6{M^{2}_{\rm Pl}}{H^2}}}} \right),
\end{equation}
and in turn, by substituting Eq. (\ref{eq72}) and relation
(\ref{eq67}) into relation (\ref{eq63}), we get
\begin{eqnarray}\label{eq73}
\frac{{{d^2}\eta }}{{d{\nu ^2}}} = &&\!\!\!\!\! E_0^2{H^2}{\left(
{1 + z}\right)^4}\Bigg[ \frac{{{d^2}\eta }}{{d{z^2}}}
+\frac{1}{{\left( {1 + z} \right)}}\times \cr
 &&\qquad \left( {3 +
\frac{{ {{\rho ^{\left( {\rm tot} \right)}} + 3{p^{\left( {\rm
tot} \right)}}} }}{{6{M^{2}_{\rm Pl}}{H^2}}}} \right)\frac{{d\eta
}}{{dz}} \Bigg].
\end{eqnarray}
Finally, by inserting Eq. (\ref{nul geo}) into Eq. (\ref{eq73}),
the null GDE corresponding to the chameleon scalar field model for
dust matters is obtained as
\begin{eqnarray}\label{null gde}
\frac{{{d^2}\eta }}{{d{z^2}}} \!\!&+&\!\! \frac{3}{{1 + z}}\left(
1 + \frac{ \rho^{\left( {\rm tot} \right)} + 3p^{\left( {\rm tot}
\right)}} {{18M_{\rm Pl}^2{H^2}}} \right)\frac{{d\eta }}{{dz}} \cr
  \!\!&+&\!\!\frac{{\left( {{\rho ^{\left( {\rm tot} \right)}} +
{p^{\left( {\rm tot} \right)}}} \right)}}{{2{{\left( {1 + z}
\right)}^2}M_{\rm Pl}^2{H^2}}}\eta  = 0.
\end{eqnarray}
However, using relations (\ref{friedman}) and (\ref{wt}),
Eq.~(\ref{null gde}) reads
\begin{equation}\label{eq75}
\frac{{{d^2}\eta }}{{d{z^2}}} + \frac{{\left( {7 + 3{w^{(\rm
tot)}}}\right)}}{{2\left( {1 + z} \right)}}\frac{{d\eta }}{{dz}} +
\frac{{3\left( {1 + {w^{(\rm tot)}}} \right)}}{{2{{\left( {1 + z}
\right)}^2}}}\eta  = 0.
\end{equation}

An analytical solution of Eq.~(\ref{eq75}), as a linear
homogeneous second-order ordinary differential equation, is
\begin{eqnarray}\label{sol}
\eta \left( z \right) = &&\!\!\!\!{C_1}{\left( {1 + z}
\right)^{\frac{1}{4}\left( { - 5 - 3{w^{\left( {\rm tot} \right)}}
- \left| {1 + 3{w^{\left( {\rm tot} \right)}}} \right|}
\right)}}\cr
 &&\!\!\!\! + {C_2}{\left( {1 + z}
\right)^{\frac{1}{4}\left( { - 5 - 3{w^{\left( {\rm tot} \right)}}
+ \left| {1 + 3{w^{\left( {\rm tot} \right)}}} \right|} \right)}},
\end{eqnarray}
where $ {C_1} $ and $ {C_2} $ are integration constants. Relation
(\ref{sol}) is the obtained result for the null vector fields in
the chameleon model that corresponds to the related one obtained
in the Brans-Dicke theory~\cite{RasouliShojai}. With appropriate
initial conditions $ \eta \left( 0 \right) = 0 $ and ${\left.
{d\eta /dz} \right|_{z = 0}} = 1 $, solution (\ref{sol}) gives
\begin{equation}
{C_1} =  - {C_2} = \frac{{ - 2}}{{\left| {1 + 3{w^{\left({\rm
tot} \right)}}} \right|}},
\end{equation}
where the derivative of the state parameter with respect to the
redshift has been assumed to be zero at the present time. We have
plotted the behavior of the deviation vector with respect to the
redshift for some range of allowed values of the total state
parameter in the top diagram of Figure~2.
%%%%%%%%%%%%%%%%%%%%%%%%%%%%%%%%%%%%%%%%%%%%%%%%%%%%%%%%%%%
\begin{figure}[h]
\includegraphics[scale=0.65]{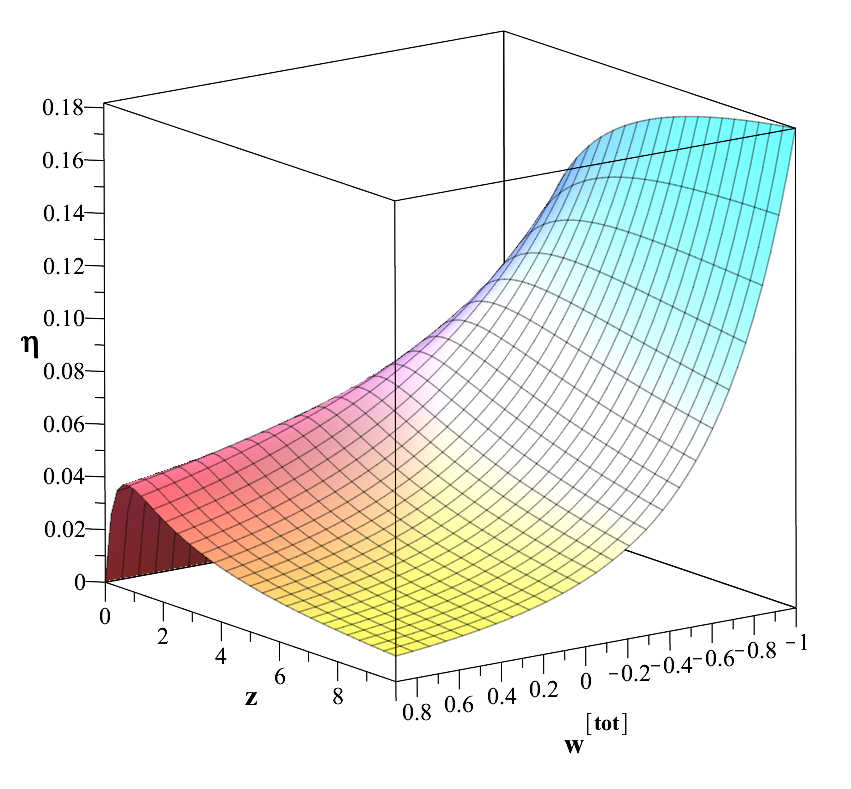}
\includegraphics[scale=0.65]{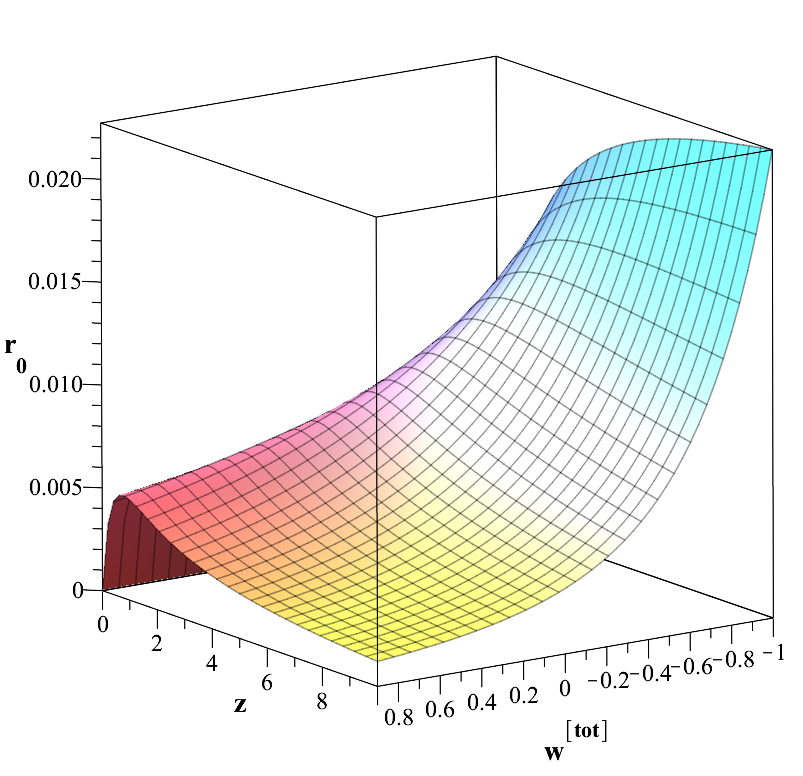}
\caption{The behavior of the deviation vector (the top diagram)
and the observer area-distance (the bottom diagram) are depicted
with respect to the redshift for null vector fields with the
initial conditions $\eta(z)|_{z=0}=0$, $d\eta(z)/dz|_{z=0}=1$ and
the parametric values $H_0\simeq 67.6$ (km/s)/Mpc.}
\end{figure}
%%%%%%%%%%%%%%%%%%%%%%%%%%%%%%%%%%%%%%%%%%%%%%%%%%%%%%%%%%%%%

Furthermore, we can indicate the observer area-distance
${r_0}\left( z \right) $ that is given by
\begin{equation}\label{r0}
{r_0}\left( z \right) := \sqrt {\left| {\frac{{d{A_0}\left( z
\right)}}{{d\Omega}_0}}\right|}  = \left| {\frac{{\eta \left( {z'}
\right)\left| {_{z' = z}} \right.}}{{d\eta \left( {z'}
\right)/dl\left| {_{z' = 0}} \right.}}} \right|,
\end{equation}
where ${A_0}$ is the area of the object and $\Omega_0$ is the
solid angle~\cite{Schneider,Ellis}. In this respect, using the
relation $\left| {d/dl} \right| = E_0^{ - 1}{\left( {1 + z} \right)^{ - 1}}d/d\nu
=H\left( {1 + z} \right)d/dz $, wherein $ dl = a\left( t \right)dr
$, while assuming that the deviation vector to be zero at $z=0$,
relation (\ref{r0}) yields
\begin{equation}\label{r00}
{r_0}\left( z \right) = \left| {\frac{{\eta \left( z
\right)}}{H_0{d\eta\left( {z'} \right)/dz'\left| {_{z' = 0}}
\right.}}} \right|.
\end{equation}
This relation represents the observed area-distance as a function
of the redshift in units of the present-day Hubble
radius~\cite{Ellis}. By inserting solution~(\ref{sol}) into
relation~(\ref{r00}), we attain the observed area-distance for the
chameleon model with dust matters, which has been depicted in the
bottom diagram of Figure~2. The comparison of the diagrams in
Figure~2 with the corresponding ones in the $ \Lambda \rm CDM $
model~\cite{Ellis}, the $ f(R) $ theory~\cite{Guarnizo}, the
Hu-Sawicki models~\cite{Dombriz}, the $ f(R,T) $
theory~\cite{zare1}, the Brans-Dicke theory~\cite{RasouliShojai}
and the space-time-matter theory~\cite{zare2} indicates that the
general behavior of the null geodesic deviation and the observer
area-distance in the chameleon model are similar to these models.
The similarity of our results to the corresponding ones in the $
\Lambda \rm CDM $ model reveals that the chameleon model remains
phenomenologically viable and can be tested with the observational
data~\cite{Ellis}.

\section{Conclusions}\label{Sec6}
We have considered the chameleon model as one of the scalar-tensor
theories, in which the chameleon scalar field non-minimally
interact with the baryonic matter field, within the homogeneous
and isotropic spatially flat FLRW background. The scalar field in
such theories (which usually plays the role of dark energy) have
been introduced in the Einstein gravitational theory to explain
the accelerated expansion of the universe. However, such a scalar
field in the chameleon model (as a chameleon field) was introduced
to remedy the problem of the EP-violation as well. On the other
hand, it would be interesting to be able to describe the evolution
of the universe with just one single scalar field from inflation
till late-time. In this regard, we had investigated the chameleon
model during inflation~\cite{saba,saba1}, wherein it has been
shown that, at the beginning of the inflation, the cosmological
constant drives the inflation and then the chameleon scalar field
plays the role of inflation~\cite{saba1}. Now, in this work, the
role of the chameleon scalar field as dark energy has been
studied.

It has been indicated that the chameleon model for dust matters
with the strong coupling and positive values of the $n$ parameter
can explain the late-time accelerated expansion of the universe.
Hence, such a model justifies dark energy with stronger
confirmation. The results not~only reveal that the strongly
coupling chameleon scalar field is viable at the late-time, but
also set some constraints on the potential of the model. Also, in
this work, we have obtained the total state parameter and shown
that the case of matter-dominated epoch causes a decelerated
evolution, and the case of the chameleon scalar field-dominated
epoch corresponds to an accelerated phase. Moreover, the analysis
shows that the inverse power-law potential remains a
model-consistent with the explanation of the universe at the
late-time.

On the other side, in order to make the investigations more
instructive, we have calculated the GDE in the chameleonic scalar
field model for the timelike and null vector fields to study the
relative acceleration of these geodesics as an effect of the
curvature of the spacetime. The case of the timelike vector fields
gives the generalized Raychaudhuri equation. The presence of the
fifth force in the chameleon model leads to the appearance of some
new terms in the GDE and the Raychaudhuri equation, which are a
direct consequence of the coupling between the chameleon scalar
field and matter field. Furthermore, we have obtained the value of
the transition redshift from the matter-dominated phase to the
late-time accelerated phase of the universe for the chameleon
model. The obtained values indicate that the transition redshift
in the chameleon model with dust matters is less than the
corresponding value of the $\Lambda \rm CDM $ model but is similar
to the $ f(R,T) $ gravity model. Hence, we conclude that the
coupling of the matter field with the scalar field or with the
geometry leads to a longer matter-dominated epoch in the evolution
of the universe.

Also, we have specified the observed area-distance of the model
through the GDE of the null vector fields. Moreover, the results
show that the general behaviors of the null deviation vector
fields and the observer area-distance in the chameleon model for
dust matters have an evolution almost similar to the other
corresponding modified gravity models. In fact, the behavior of
almost all models, for the deviation evolution, is similar to the
$\Lambda \rm CDM $ model for small values of $ z $, as expected.
That is, their results remain like a cosmological constant with
small corrections to GR.

The relations of the area-distance (\ref{r0}) and (\ref{r00}) can
be used for the angular size-redshift relation derived from the
Sunyaev-Zel'dovich effect x-ray technique~\cite{Bonamente, Chen},
and also to compact the radio sources as cosmic rulers~\cite{Lima,
Jackson}. Moreover, by considering the relation between the area-
and the luminosity-distances (see, e.g. Ref.~\cite{Matravers}), it
is also possible to extend the investigations in the chameleon
model with the data obtained from the observations of type Ia
supernovae~\cite{Suzuki, Campbell}. In addition to the observed
area-distance, the geodesic deviation equation can be used to
study the effect of the generalized tidal forces, which could lead
to the possibility of observationally testing the model through
the observational effects of tides due to an extended mass
distribution, for more details see Refs.~\cite{Porciani,
Harko-GDE}.

 %%%%%%%%%%%%%%%%%%%%%%%%%%%%%%%%%%%%%%%%%%%%%%%%%%%%%%%%%%%%%%%%%%%%%%%

\end{document}